\documentstyle[12pt]{article}
\begin{document}
\begin{flushright}
DPNU-94-57\\
hep-th/9502043\\
Feb. 1995
\end{flushright}
\vspace{5pt}
\setlength{\baselineskip}{0.25in}
\renewcommand{\thefootnote}{\fnsymbol{footnote}}
\begin{center}
{\Large{\bf NAMBU-GOLDSTONE BOSON
ON THE LIGHT-FRONT
\footnote
{To appear in the proceedings of the
13th Symposium on Theoretical Physics,
Mt. Sorak, Korea,\\
\hspace*{15pt}
from 27 June to 2 July, 1994}}}
\vspace{30pt}
\noindent

Yoonbai Kim\footnote[4]{E-mail address:
yoonbai@eken.phys.nagoya-u.ac.jp},
Sho Tsujimaru\footnote[2]{E-mail address:
sho@ins.u-tokyo.ac.jp} and
Koichi Yamawaki\hspace{2pt}\footnote[3]
{E-mail address:
yamawaki@eken.phys.nagoya-u.ac.jp}\\
\vspace{10pt}
\noindent

Department of Physics, Nagoya University,
Nagoya 464-01, Japan\\
$\mathop{}^{\dagger}$Institute for Nuclear
Study, University of Tokyo, Tanashi-shi,
Tokyo 188, Japan
\end{center}
\vspace{5mm}

\begin{abstract}
Spontaneous breakdown of the continuous
symmetry is studied in the
framework of discretized light-front
quantization. We consider linear
sigma model in 3+1 dimension and
show that the careful treatment of
zero modes together with the
regularization of the theory by
introducing NG boson mass leads to the
correct description of Nambu-Goldstone
phase on the light-front.
\end{abstract}

\vspace{5mm}

Recently the light-front (LF) quantization
with a Tamm-Dancoff
truncation has attracted much attention
as a promising method for
solving QCD and other strong coupling
theories and indeed it
describes the bound state spectra
successfully in various field
theoretical models in (1+1)
dimensions \cite{CHA}. However, there remain
difficulties in applying  the
 present-stage formulation directly to
(3+1) dimensional gauge theories
which are working tools of modern
particle physics.
One of them is to understand the
phenomenon of spontaneous
symmetry breaking (SSB) on the LF
and it may provide a cornerstone
for the nonperturbative LF QCD. For the discrete
symmetry, several authors discussed
the possibility of SSB in
(1+1) dimensional scalar models
and argue that the solution of the
zero-mode constraints\cite{MY} may realize
the broken vacuum \cite{HKW}.

In this report  we  address ourselves to
the problem of SSB of the continuous
symmetry on LF in the
linear sigma model.
Based
on the discretized light front
quantization (DLFQ) \cite{MY,PB}, we
clarify how the Nambu-Goldstone (NG) phase
is realized through the
careful treatment of zero-mode constraints \cite{MY}
together with an
infrared regularization by explicit
symmetry-breaking mass of
NG boson $m_{\pi}$\cite{KTY}.
The NG-boson zero mode, when integrated
over the transverse space, must behave
as singular $\sim 1/m_{\pi}^2$ 
in the symmetric limit $m_{\pi}^2 \rightarrow 0$.
This result is actually valid model-independently,
though we demonstrate
it in a specific model field theory with 
explicit use of the zero-mode constraints.

 In what follows we shall first show that
naive use of the zero mode constraints leads
to the inconsistent
result that the LF formalism allows neither
the coupling of NG boson
nor the current vertex associated with NG boson.
Let us consider the $O(2)$ linear sigma
model which 
is defined by the Lagrangian
\begin{equation}\label{lag}
{\cal L}=\partial_{+}\sigma
\partial_{-}\sigma+\partial_{+}\pi
\partial_{-}\pi-\frac{1}{2}(\partial_{\bot}\sigma)^2
-\frac{1}{2}
(\partial_{\bot}\pi)^2-\frac{1}{2}\mu^2
(\sigma^2+\pi^2)-
\frac{\lambda}{4}(\sigma^2+\pi^2)^2,
\end{equation}
where $x^{+}=\frac{1}{\sqrt{2}}(x^{0}+x^{3})$
plays the role of LF time,
$x^{-}=\frac{1}{\sqrt{2}}(x^{0}-x^{3})$
is the LF spatial coordinate
($-L\leq x^{-}\leq L$) and $x^{\bot}$
denotes transverse coordinate $(x^{1}, x^{2})$.
Here we assume the periodic boundary
condition to the fields.
The Lagrangian in Eq.(\ref{lag}) is
invariant under $O(2)$
transformation, and then there is
a conserved current $J_{\mu}
=\partial_{\mu}\sigma\pi-\partial_{\mu}\pi\sigma$.

In the framework of the canonical DLFQ,
it is convenient to decompose the field
degrees $\pi$ (or $\sigma$) into
the oscillating modes ($P^{+}\neq0$),
$\varphi_{\pi}$ (or $\varphi_{\sigma}$),
described by the following
commutation relations which are the
same as those of free fields \cite{MY}:
\begin{equation}\label{comm}
[\varphi_i(x),\varphi_j(y)]=-{i \over 4}
\Bigl\{\epsilon(x^--y^-)
-{x^--y^- \over L}\Bigr\}
\delta_{ij}\delta^{(2)}(x^{\bot}-y^{\bot}),
\end{equation}
where $i, j$ denote $\pi$ or $\sigma$,
and the zero modes ($P^{+}=0$),
$\pi_{0}$ (or $\sigma_{0}$), depicted
by the solutions of
zero-mode constraints \cite{MY}:
\begin{eqnarray}
\chi_{\pi}&\equiv&\displaystyle{\frac 1{2L}
\int^L_{-L}dx^-}
\Bigl[(\mu^2-\partial_{\bot}^2)\pi
+\lambda\pi(\pi^2+\sigma^2)\Bigr]
\approx 0
\label{ze1}\\
\chi_{\sigma}&\equiv&\displaystyle{\frac{1}{2L}
\int^L_{-L}dx^-}
\Bigl[(\mu^2-\partial_{\bot}^2)\sigma+
\lambda\sigma(\pi^2+\sigma^2)
\Bigr]\approx 0,
\label{ze2}
\end{eqnarray}
which are the consequences of DLFQ with
periodic boundary condition
\begin{equation}
-\frac{1}{2L}\int^L_{-L}dx^-2
\partial_{+}\partial_{-}\pi=-\frac{1}{2L} \int^L_{-L}
dx^-2 \partial_{+}\partial_{-}\sigma=0.
\end{equation}

To solve explicitly the zero-mode constraints 
Eq.(\ref{ze1}) and Eq.(\ref{ze2})
(or the modified ones after
introducing explicit symmetry
breaking term to be discussed later),
we divide them into two parts:
The classical constant pieces whose
solution is the classical vacuum solution 
chosen as $v_{\pi}=0$ and $v_{\sigma}
\equiv v=\sqrt{-\frac{\mu^{2}}{\lambda}}
$, and the operator part expanded as a
perturbative series in $\lambda$:
\begin{equation}\label{opze}
\omega_i=\sum_{k=1}\lambda^k \omega_i^{(k)},
\end{equation}
where $\omega_{\pi}\equiv \pi_0-v_\pi$,
$\omega_{\sigma}\equiv 
\sigma-v_{\sigma}$.  
Each  $\omega^{(k)}_{i}$ is determined
recursively by inserting
Eq.(\ref{opze}) into the zero-mode
constraints in Eq.(\ref{ze1}) and
Eq.(\ref{ze2}). A straightforward
calculation leads to the first-order
perturbation:
\begin{eqnarray}
\partial^{2}_{\bot}\omega^{(1)}_{\pi}
&=&\frac{1}{2L}\int^{L}_{-L}dx^{-}
\Bigl(\varphi_{\pi}^{3}+\varphi_{\pi}
\varphi^{2}_{\sigma}+2v
\varphi_{\pi}\varphi_{\sigma}\Bigr),
\label{fir1}\\
(-m_{\sigma}^{2}+\partial^{2}_{\bot})
\omega^{(1)}_{\sigma}&=&
\frac{1}{2L}\int^{L}_{-L}dx^{-}
\Bigl(\varphi^{3}_{\sigma}+
\varphi^{2}_{\pi}\varphi_{\sigma}
+3v\varphi^{2}_{\sigma}
+\varphi^{2}_{\pi}\Bigr),
\label{fir2}
\end{eqnarray}
where $m_{\sigma}^2=2\lambda v^2$.
Note that we assume the Weyl ordering
between the zero modes and
the non-zero modes throughout
this report but omit it
in the expressions of formulas for simplicity.
Actually, as far as the tree level
operator solution (7) and (8) is concerned,
it is determined independently of the ordering.
The Hilbert space of our system is
constructed without zero mode
and the vacuum $\vert 0\rangle$
is trivial, since the LF momentum
$P^{+}$ is positive definite
without referring to the dynamics.

The zero-mode part of the LF charge including the
NG-boson pole term is removed by
integration on the LF and
the periodic boundary condition of the fields,
so that the LF charge
is expressed only by the non-zero modes:
\begin{equation}\label{lfch}
Q=\lim_{L\rightarrow \infty}
\int^{L}_{-L}dx^-d^2x^{\bot}
(\partial_{-}\varphi_{\sigma}
\varphi_{\pi}-\partial_{-}
\varphi_{\pi}\varphi_{\sigma}).
\end{equation}
The LF charge in Eq.(\ref{lfch}) always
annihilates the vacuum
$\vert 0\rangle$ due to the conservation of the LF
momentum $P^{+}$ and thereby
it seems to be a well-defined quantity
independently of the detailed information
of the phase of the system.
This is also consistent with the
explicit computation of the
commutators: 
$\langle [Q, \varphi_{\sigma}]\rangle =-i
\langle\varphi_{\pi}\rangle=0$ and 
$\langle [Q,\varphi_{\pi}]\rangle =i\langle
\varphi_{\sigma}\rangle=0$,\footnote{
By explicit calculation with a careful
treatment of the zero modes
contribution, we can also show that
$\langle [Q, \sigma]\rangle =
\langle [Q,\pi]\rangle =0$.
}
 which are contrasted to those in the usual
equal-time case 
 where the spontaneously broken charge
does not 
annihilate the vacuum:
$\langle [Q^{\rm et},\sigma]\rangle
=-i\langle \pi\rangle=0,
\langle [Q^{\rm et},\pi]\rangle=i\langle
\sigma\rangle\ne 0$.

Using Eqs.(\ref{fir1}), (\ref{fir2})
and neglecting the trivial
divergence from the operator ordering,
we can check the conservation
of the LF charge to the leading order:
\begin{equation}
[Q, P^-]=i\int dx^-d^2 x^{\bot}
(v\chi_\pi+\omega_{\sigma}
\chi_{\pi}-\omega_{\pi}\chi_{\sigma})=0.
\end{equation}
The vacuum annihilation and the conservation
of the LF charge do not hold
simultaneously in the NG phase in the
 conventional equal-time
quantization, so that 
one may easily expect some inconsistency
in the NG phase.

To clarify the underlying inconsistency,
let us compute explicitly
the $\sigma\rightarrow\pi\pi$ vertex at
the tree level.
Based on the LSZ reduction formula, we have
\begin{eqnarray}
\lefteqn{\langle \pi\pi(q)\vert\sigma\rangle
\equiv i \int d^4 xe^{iqx} \langle
\pi\vert \Box \pi(x)
\vert \sigma\rangle} \nonumber \\
&=&i(2\pi)^4 \delta(p_{\sigma}^{-}
-p_{\pi}^{-}-q^-)\delta(p_{\sigma}^{+}
-p_{\pi}^{+}-q^+)\delta(p_{\sigma}^{\bot}-
p_{\pi}^{\bot}-q^{\bot})\langle\pi
\vert j_{\pi}(0)\vert\sigma\rangle ,
\end{eqnarray}
where
$q_{\mu}=p_{\mu}^{\sigma}-p_{\mu}^{\pi}$ is
the momentum of NG boson
and $j_{\pi}(x)=\Box \pi(x)=(2\partial_{+}
\partial_{-}-\partial_{\bot}^2)\pi$ is
the source function of the NG boson which
is given in our model as
$j_{\pi}=-\lambda(\pi^{3}+\pi\sigma^{'2}
+2v\pi\sigma^{'})$,
with $\sigma^{\prime}=\sigma-v$.
Taking the collinear momentum frame,
$q^+=q^{\bot}=0$
and $q^{-}\ne 0$ for the
emission vertex of the exactly massless
NG boson with $q^2=0$, we find
that the NG boson emission vertex does
vanish as follows:
\begin{eqnarray}
\lefteqn{(2\pi)^3\delta(p_{\sigma}^{+}
-p_{\pi}^{+})\delta^{(2)}
(p_{\sigma}^{\bot}-p_{\pi}^{\bot})
\langle\pi\vert j_{\pi}(x^+,0)\vert\sigma\rangle}\nonumber\\
&=&\lim_{L\rightarrow \infty}\int_{-L}^{L}
dx^-d^{2}x^{\bot} \langle\pi
\vert(2\partial_{+}\partial_{-}-
\partial_{\bot}^2)\pi\vert\sigma\rangle\label{res1}\\
&=&\int d^{2}x^{\bot}\lim_{L\rightarrow\infty}
\langle\pi\vert
\Bigl(\int^{L}_{-L}dx^{-}2\partial_{+}
\partial_{-}\pi\Bigr)\vert\sigma\rangle=0.\nonumber
\end{eqnarray}
On the other hand, by use of the explicit
interaction term
$j_{\pi}=-\lambda(\pi^{3}
+\pi\sigma^{'2}+2v\pi\sigma^{'})$, the NG
boson vertex can also be calculated
at the tree level as
\begin{equation}\label{spp}
\lim_{L\rightarrow \infty}\int_{-L}^{L}
dx^{-}d^{2}x^{\bot}\langle\pi\vert
j_{\pi}(x)\vert\sigma\rangle=-2\lambda v
(2\pi)^{3}\delta(q^{+})\delta(q^{\bot}),
\end{equation}
which is in agreement with the result
in equal-time formalism.
If two calculations (12) and (13)
were to be compatible, $v=0$ would be
concluded for the interacting
theory ($\lambda\ne 0$), and thus
the NG boson is completely decoupled on the LF.

According to
the Goldberger-Treiman relation in
the context of conventional
canonical quantization, the NG boson
emission vertex is related to
the current vertex and then we should
check whether the above symptom
is conveyed to the current vertex or not.
Suppose that the NG phase
is realized, then the current contains
the NG-boson term:
$J_{\mu}=-v\partial_{\mu}
\pi+\widehat J_{\mu}$,
 where the non-pole term is given by
$\widehat J_{\mu}=
\pi\partial_{\mu}\sigma^{\prime}-\partial_{\mu}
\pi \sigma^{\prime}$.
of the NG boson.
Here we can easily notice that
the non-pole charge $\widehat Q=
\lim_{L\rightarrow\infty}
\int_{-L}^{L}dx^- d^2 x^{\bot}\widehat J^{+}$
is equal to $Q$, $\widehat{Q}=Q$,
that is, $d\widehat Q /dx^{+}
=[\widehat Q ,P^{-}]=0$, which
implies  that
the current vertex $\langle\pi
\vert\hat{J}^{+}(0)\vert\sigma
\rangle$ also vanishes:
\begin{eqnarray}
\lefteqn{0=\lim_{L\rightarrow \infty}
\langle\pi\vert \int_{-L}^{L}
dx^- d^2 x^{\bot}\partial_{\mu}
\widehat J^{\mu}(x)\vert\sigma
\rangle_{x^+=0}}\nonumber\\
&=&-i(2\pi)^3\delta(q^+)\delta^{(2)}
(q^{\bot})\displaystyle{
\frac{m_{\sigma}^2-m_{\pi}^2}{2p_{\sigma}^+}}
\langle\pi\vert
\widehat J^+(0)\vert\sigma\rangle,\label{res2}
\end{eqnarray}
as far as $m^{2}_{\sigma}\neq m^{2}_{\pi}$.
Now we arrive at a``no-go
theorem" that the NG boson cannot exist on the LF. 
Thus it is too naive to expect\cite{HKW}
that the NG phase can be realized by
simply solving the zero-mode constraints.

As is easily seen
from its derivation, this ``no-go theorem''
indeedreflects a genuine nature of LF coordinate
(first-order form of
$\Box=2\partial_{+}\partial_{-}-\partial^2_{\bot}$
in $\partial_{-}$) and the periodic boundary
condition, and hence holds model-independently.
In fact, we can derive (\ref{res1}) and (\ref{res2}),
based on the LSZ reduction formula and
the current expression
 containing the NG boson pole term:
$J_{\mu}=-f_{\pi}\partial_{\mu}\pi
+\widehat J_{\mu}$, where 
 $\pi$ is the interpolating field of
the NG boson and $f_{\pi}$
the decay constant ($=v$ 
in the linear sigma model).

How can the NG boson live on the LF?
Here we propose 
 a regularization by introducing the explicit
symmetry breaking through the NG boson mass
$m_{\pi}^{2}$ and recovering it in the
massless limit $m_{\pi}^{2}
\rightarrow 0$ \cite{KTY}. To be specific
we add an explicit symmetry breaking
term ${\cal L}_{reg}=c\sigma$ to the
 Lagrangian in Eq.(\ref{lag})and then
the NG boson has mass $m^{2}_{\pi}=
\mu^{2}+\lambda v^{2}$ and the
$\sigma$-meson mass shifts to
$m_{\sigma}^{2}=\mu^{2}+3\lambda v^{2}$,
where the vacuum expectation value $v$
is determined by
$\mu^{2}v+\lambda v^{3}=c$.
Since the symmetry is explicitly broken,
the current is not conserved
and the current vertex is
changed, i.e., $\partial_{\mu}J^{\mu}(x)=
vm_{\pi}^2\pi(x)$ and $\partial_{\mu}\widehat J^{\mu}(x)=
v(\Box+m_{\pi}^2)\pi(x)=vj_{\pi}(x)$, but, at classical
level, the symmetry is restored in the limit of massless
NG boson.
The zero-mode constraints are modified accordingly
and a series of solutions are evaluated
in the perturbation on  the
coupling constant $\lambda$.
The first order solution for the $\sigma$
part is formally the same as Eq.(8) while the
NG boson part has a mass term  as
\begin{equation}\label{pio}
(-m_{\pi}^{2}+\partial^{2}_{\bot}
)\omega^{(1)}_{\pi}
=\frac{1}{2L}\int^{L}_{-L}dx^{-}
(\varphi^{3}_{\pi}
+\varphi_{\pi}
\varphi^{2}_{\sigma}+2v\varphi_{\pi}
\varphi_{\sigma}).
\end{equation}

In spite of the addition of tiny
NG boson mass term to the Lagrangian,
the behavior of NG boson zero mode
shows a drastic
change. Since the R.H.S. of  Eq.(\ref{pio})
is finite in the limit of zero
 NG boson mass, the NG boson zero mode
integrated over $x^{\bot}$
 space does become singular:
\begin{equation}
\int d^{2}\!x^{\bot}\omega_{\pi}^{(1)}\sim
\frac{1}{m^{2}_{\pi}}\hspace{10mm}
(m_{\pi}^{2}\rightarrow0).
\end{equation}
Using this zero mode in
Eq.(\ref{pio}), we
achieve the recovery of both the non-zero
NG boson vertex
and the current vertex:
\begin{eqnarray}\label{cver}
\lim_{L\rightarrow \infty}
\langle\pi\vert \int_{-L}^{L} dx^-d^2x^{\bot}
\partial_{\mu}\widehat J^{\mu}(x)
\vert\sigma\rangle
&=&\lim_{L\rightarrow \infty}
\int_{-L}^{L}dx^{-}d^{2}x^{\bot}
\langle \pi \vert vj_{\pi}(x)\vert
\sigma\rangle \nonumber \\
&=&v m^{2}_{\pi}\langle\pi\vert
\lim_{L\rightarrow\infty}
\int dx^{-}d^{2}\!x^{\bot}\pi(x)
\vert\sigma\rangle \nonumber \\
&=&-2\lambda v^{2}(2\pi)^{3}\delta(q^{+})
\delta^{(2)}(q^{\bot}),
\end{eqnarray}
even when we take the limit of zero NG boson mass.

Though the vacuum annihilation property
of the LF charge is not affected by the
regularization,
the non-zero current vertex implies
that the LF charge is now non-conserving:
\begin{equation}
[Q, P^-]=ivm_{\pi}^2
\lim_{L\rightarrow \infty}
\int_{-L}^{L} dx^{-}d^2x^{\bot}
\pi\stackrel{m_{\pi}^2\rightarrow
0}{\neq} 0.
\end{equation}
This also can be checked by direct
computation of the commutator of $Q$ 
with $P^{-}$.

A way to confirm the validity of the
introduction of regularization
on the LF is to envisage
the momentum space expression of the
relation $\partial^{\mu} J_{\mu}
=vm_{\pi}^2\pi$(operator relation of PCAC):
\begin{equation}\label{mom}
\displaystyle{
\frac{m_{\pi}^2 v j_{\pi}(q)}{m_{\pi}^2-q^2}
(=\partial^{\mu}J_{\mu}(q))=
\frac{q^2 v j_{\pi}(q)}{m_{\pi}^2-q^2}
+\partial^{\mu}\widehat J_{\mu}(q). }
\end{equation}
Above all it is necessary to summarize
what we have done when we reached
the false ``no go theorem''. The L.H.S.
of Eq.(\ref{mom})
was set to be zero, because we took an
account of a situation of massless
NG boson where the current is conserved,
while the first term in the R.H.S. of
Eq.(\ref{mom}) is also dropped
due to the periodic boundary condition
or the zero-mode constraint
on the LF. It is obvious that the
above procedure
is equivalent to bringing up a nonsense
$\lim_{m_{\pi}^{2},\,q^{2}\rightarrow 0}
(\frac{m^{2}_{\pi}-q^{2}}{m^{2}_{\pi}-q^{2}})
=0$ as far as  $vj_{\pi}(q)\neq 0$ (NG phase).
In the conventional equal-time quantization 
the same correct results can be obtained
independently of the order of the limits
$q^{2}\rightarrow 0$ and
$m_{\pi}^{2}\rightarrow 0$, while in
the LF with $q^{+}=q^{\bot}=0$
we always take $q^{2}\equiv 0$ first
and then $m_{\pi}^{2}=0$
must be regarded as a limit
$m_{\pi}^{2}\rightarrow0$\cite{CHKT}.
Therefore the L.H.S. of Eq.(\ref{mom})
does not vanish and hence
the LF charge is not conserved in the NG phase.
Since Eq.(\ref{mom}) holds
model-independently (PCAC), the above
arguments are also valid in the general framework.

Our treatment of the zero mode in the canonical
DLFQ is quite different from that proposed recently
by Wilson et al.\cite{WW} who
eliminate the zero mode in the continuum theory
without specifying the
boundary condition at $x^{-}=\pm\infty$.
Although our conclusion
on the trivial vacuum and the non-conservation
of the LF chargeappears to be consistent with
that in Ref.\cite{WW},
the relationship between these two approaches
are not clear at the moment.

\vspace{5mm}

{\Large{\bf\noindent Acknowledgements}}

\vspace{5mm}

We would like to thank T. Kugo, Y. Ohnuki and
I. Tsutsui
for useful discussions. Y.K. thanks the members of
Center for Theoretical Physics in Seoul National
University for valuable discussions
and their hospitality.
This work was supported in part by a
Grant-in-Aid for Scientific Research from the
Ministry of Education,
Science and Culture (No.05640339), the
Ishida Foundation and the Sumitomo
Foundation (K.Y.), and 
also by JSPS under No.93033 (Y.K.).
A part of this work was done while K.Y. was
staying at Institute for Theoretical
 Physics at U.C. Santa Barbara, being
supported in part by the Yoshida
  Foundation for Science and Technology
 and by the U.S. National Science Foundation
under Grant No.PHY89-04035.

\end{document}